\documentclass[referee]{raa}            

\usepackage{graphicx}             
\usepackage{times}
\usepackage{natbib}
\usepackage{appendix}
\usepackage{amsmath}
\usepackage{amssymb}
\usepackage{epsfig}
\usepackage{rotating}
\usepackage{color}
\usepackage{longtable}
\usepackage{pdflscape,lscape}
\usepackage{float}
\usepackage{placeins}
\usepackage{booktabs}
\usepackage{upgreek}
\usepackage{url}
\usepackage[a4paper=true]{hyperref}
\hypersetup{colorlinks = true, linkcolor = green, anchorcolor = red, citecolor = blue, filecolor = red, pagecolor = red, urlcolor = red}
\begin{document}

   \title{The evolutionary status of chemically peculiar eclipsing binary star DV Boo
}

   \volnopage{Vol.0 (20xx) No.0, 000--000}      
   \setcounter{page}{1}          

   \author{F. Kahraman Ali\c{c}avu\c{s}
      \inst{1,2}
   \and F. Ali\c{c}avu\c{s}
      \inst{1}
   }

   \institute{\c{C}anakkale Onsekiz Mart University, 
Faculty of Sciences and Arts, Physics Department, 17100, \c{C}anakkale, Turkey; {\it filizkahraman01@gmail.com}\\
        \and
             Nicolaus Copernicus Astronomical Center, Bartycka 18, PL-00-716 Warsaw, Poland\\
\vs\no
   {\small Received~~20xx month day; accepted~~20xx~~month day}}

\abstract{Eclipsing binary systems are unique stellar objects to examine and understand the stellar evolution and the formation. Thanks to these systems, the fundamental stellar parameters (mass, radius) can be obtained very precisely. The existence of metallic-line (Am) stars in binaries is noticeably common. However, the known number of Am stars in eclipsing binaries is less. The Am stars in eclipsing binaries are extremely useful to deeply investigate the properties of Am stars, as the eclipsing binaries are the only tool to directly derive the fundamental stellar parameters. Additionally, the atmospheric parameters and the metallicities of the binary components could be obtained by a detailed spectroscopic study. Therefore, in this study, we present a comprehensive photometric and spectroscopic analysis of the eclipsing binary system DV\,Boo which has a possible Am component. The fundamental stellar parameters were determined by the analysis of radial velocity and photometric light curves. The atmospheric parameters of both binary components of DV\,Boo were derived using the disentangled spectra. The chemical abundance analysis was carried out as well. As a result, we showed that the primary component illustrates a typical Am star chemical abundance distribution. The fundamental stellar parameters of the binary components were also obtained with an accuracy of $<$1\% for masses and $<$3\% for radii. The evolutionary status of DV\,Boo was examined using the precisely obtained stellar parameters. The age of the system was found to be 1.00\,$\pm$\,0.08\,Gyr.   
}

   \authorrunning{Kahraman Ali\c{c}avu\c{s} \& Ali\c{c}avu\c{s}}            
   \titlerunning{Chemically peculiar eclipsing binary star DV Boo}  

   \maketitle
   
\section{Introduction}           
\label{sect:intro}
A significant amount of stars is a member of binary or 
multiple systems (\citealt{2014MNRAS.443.3022A, 2011IAUS..272..474S}). These objects, in particularly the eclipsing binary systems, are unique 
tools to comprehend information about the formation and the evolution of stars. Eclipsing binary systems
provide a direct measurement of the fundamental 
stellar parameters (e.g. mass, radii) with a good accuracy (\citealt{2010A&ARv..18...67T, 2013A&A...557A.119S}). These 
fundamental parameters are significant for a deep investigation of a star. To obtain the precise fundamental stellar 
parameters, both photometric and spectroscopic data are needed. The radial velocity analysis supplies the 
determination of the exact orbital parameters, while the light curve analysis provides the orbital inclination and 
the radii of the stars relative to the semi-major axis. As a result of both analyses, precise fundamental stellar parameters such as mass ($M$) could be obtained. The $M$ is the most important parameter which determines the life of a star. Therefore, an accurate determination of $M$ is essential. 
In addition to $M$, the metallicity also affects the life of a star. Hence, for a comprehensive investigation of 
the binary evolution both parameters should be obtained. The double-lined eclipsing binary systems provide an opportunity to derive these parameters. 
While the precise $M$ can be obtained from the analysis of radial velocity and light curves, the metallicity can be determined by using special approaches like spectral disentangling method (\citealt{1994A&A...281..286S}). 

The binary systems are thought to form in the same interstellar area and hence the component 
stars in the binary system should have the same chemical abundance pattern. However, the recent detailed studies about 
eclipsing binary stars showed that components in a binary system could 
have different chemical structure (e.g. \citealt{2018A&A...615A..36P, 2017JASS...34...75J}). This result is very important to understand the binary evolution. 
The increased number of such samples will be effective. Therefore, in this study, we present a detailed analysis of a suspected metallic-line eclipsing binary star, DV\,Boo. 

DV\,Boo (HD\,126031, V\,=\,7$^{m}$.54) was classified as an eclipsing binary system by the \textit{Hipparcos} (\citealt{1997ESASP1200.....E}). The star was defined as a metallic-line star by \cite{1988PASP..100.1084B}. \cite{1999A&AS..137..451G} gave a spectral classification 
of A3kA7hF5m which points out a chemically peculiar star. However, these spectral classifications could be 
suspected because the double-lined feature of DV\,Boo was later 
discovered by \cite{2004MNRAS.352..708C}. \cite{2004MNRAS.352..708C} analysed medium resolution 
(R$\sim$20\,000, 42\,000) spectroscopic data to obtain the orbital
parameters of the binary system. Additionally, they presented the analysis of the \textit{Hipparcos} light 
curve to derive the fundamental stellar parameters. However, because of the 
low-quality photometric data and less number of the measured radial 
velocities for secondary component (low-mass) these parameters could not be determined sensitively. DV\,Boo is also given in a list of candidate 
pulsating eclipsing binary star (\citealt{2006MNRAS.370.2013S}). However, the pulsation 
nature (if exist) of the star has not been discovered. 

In this study, we introduce a detailed analysis of DV\,Boo which is believed to show 
chemically peculiar characteristic. The star also has enough and good spectroscopic and 
photometric public data to investigate the star deeply. A detailed analysis of the star will 
allow us to obtain the atmospheric chemical structure of both binary components in DV\,Boo. 

The paper is organized as follows. The information about the used spectroscopic and photometric data is given 
in Sect.\,2. The analysis of the radial velocity measurements is presented in Sect.\,3. We introduced a 
comprehensive spectral analysis in Sect.\,4. The light curve analysis is given in Sect.\,5. Discussion and conclusions 
are presented in Sect.\,6.


\section{observational data}
\label{sect:Obs}

In this study, we used the public photometric and spectroscopic data of DV\,Boo. 
The system has photometric data in two different archives. First is the All Sky Automated Survey (ASAS) archive
\footnote{http://www.astrouw.edu.pl/asas} (\citealt{2002AcA....52..397P}). ASAS is a 
low budget project which aims to detect and identify variable stars. The project provides data taken with $V$- and $I$-bands. 
However, DV\,Boo has only $V$-band data in the ASAS archive. The second archive is the Kamogata/Kiso/Kyoto wide-field survey (KWS) 
archive\footnote{http://kws.cetus-net.org/$\sim$maehara/VSdata.py} (\citealt{2014SSIC..3.....119M}). This survey provides $B$-, $V$- and $Ic$-band data. However, DV\,Boo only has usable $V$- and $Ic$-band data in the archive. These available photometric data were gathered and the scattering points 
beyond the 3$\sigma$ level were removed for the light curve analysis.  

The spectroscopic data of DV\,Boo is available in ELODIE\footnote{http://atlas.obs-hp.fr/elodie/} and European Southern Observatory (ESO) \footnote{http://archive.eso.org} archives. In the ELODIE archive, there are twelve spectra of DV\,Boo. The ELODIE is an \'{e}chelle spectrograph which was mounted at the 1.93-m telescope at the Observatoire de Haute Provence (France). The spectrograph
provides spectra with a resolving power of $\sim$42000 and with a wavelength range 
of 3850\,$-$\,6800\,{\AA} (\citealt{2004PASP..116..693M}). The ELODIE spectra are 
served after an automatic data reduction by the 
dedicated pipeline. All available reduced ELODIE spectra of DV\,Boo were 
taken to use in this study. 

The ESO archive provides the data taken from the ESO instruments at La Silla Paranal observatory. 
In this archive, there are six FEROS and eight HARPS spectra of DV\,Boo. FEROS and HARPS are 
\'{e}chelle spectrographs and they are attached to the 2.2-m and the 3.6-m telescopes, respectively. 
FEROS has a resolving power of 48000 and its spectral range is approximately 
between 3500 to 9200\,{\AA} 
(\citealt{1999Msngr..95....8K}). HARPS has a higher resolving power (85000) and 
it supplies spectra in a wavelength range of about 3780\,$-$\,6900\,{\AA} 
(\citealt{2003Msngr.114...20M}). 
The spectra of FEROS and HARPS were reduced and calibrated by their 
dedicated pipelines. 
All available FEROS and HARPS spectra for DV\,Boo were used in this study. 

The collected reduced spectra were manually normalized using the 
NOAO/IRAF\footnote{http://iraf.noao.edu/} \textit{continuum} task. 
Each spectrum was normalized separately and the continuum level was controlled with a synthetic spectrum 
that was generated using approximate atmospheric parameters. The information for the used spectroscopic data 
is given in Table\,\ref{table1}. The average signal-to-noise 
(S/N) ratio of all spectroscopic data is $\sim$95. 

\begin{table}
\begin{center}
\caption[]{Information for the Spectroscopic Data. The Last Column Represents the Range of the Signal-to-noise (S/N) Ratio. }\label{table1}


 \begin{tabular}{lccccc}
  \hline\noalign{\smallskip}
 Instruments & Resolving  &Observation  & Number of         &  S/N           \\
             &  power     & dates        & Spectra           &  Range \\
  \hline\noalign{\smallskip}
ELODIE        & 42000      & 2001\,$-$\,2002 & 12  &  45\,$-$\,90 \\
FEROS         & 48000      & 2016            & 6   &  100\,$-$\,135 \\
HARPS         & 85000      & 2009\,$-$\,2013 & 8   &  80\,$-$\,140 \\
  \noalign{\smallskip}\hline
\end{tabular}
\end{center}
\end{table}


   \begin{figure}
   \centering
   \includegraphics[width=12cm, angle=0]{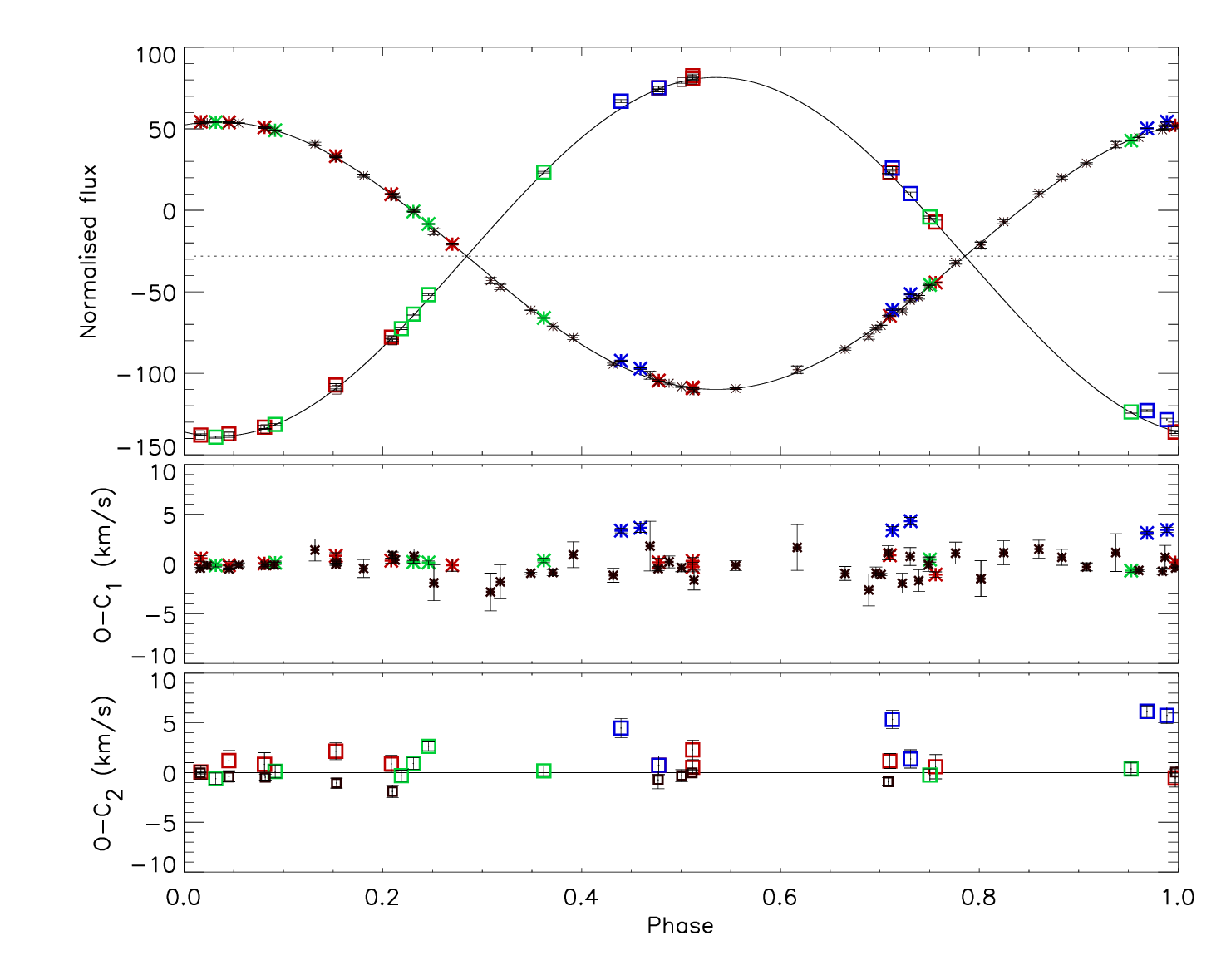}
   \caption{Theoretical fits (solid lines) to the radial velocity measurements of the primary (stars) and the secondary (squares) components are given in the upper panel. The red, blue, green, and black symbols are illustrated the ELODIE, FEROS, HARPS and the literature values (\citealt{2004MNRAS.352..708C}), respectively. Dashed line in the upper panel shows the V$\gamma$ level. The residuals from the fits are shown in the lower panel. 
The subscripts ``1'' and ``2'' represent the primary and the secondary components, respectively.}
   \label{Fig1}
   \end{figure}

\section{Radial velocity measurements}
\label{sect:rv}
A binary system is defined with its orbital parameters such as inclination $i$, 
eccentricity $e$, semi-major axis $a$ and argument of periastron $w$. To obtain the accurate orbital parameters of an eclipsing binary system, the radial velocity measurements spread over the entire orbital phases are required. 

The radial velocity ($v_{r}$) measurements of DV\,Boo were obtained by using 
the available public spectra. The IRAF, FXCOR task was used in the measurements. In this analysis, a radial velocity standard star, which has spectra from the same instruments with DV\,Boo, was taken into account. HD\,693 was chosen as a template star. We calculated the $v_{r}$ value of HD\,693 as 15.18\,km\,s$^{-1}$ on average which is similar to the literature value ($v_{r}$\,=\,14.81\,km\,s$^{-1}$, \citealt{2010A&A...521A..12M}). As a result, the $v_{r}$ values of DV\,Boo were obtained and they are listed in Table\,\ref{rvs}.

\begin{table}
\begin{center}
\caption[]{The Results of the Radial Velocity Analysis. $^{a}$ Represents the Fixed Parameters.}\label{table2}
 \begin{tabular}{lc}
  \hline\noalign{\smallskip}
  Parameters  & Value  \\
  \hline\noalign{\smallskip}
$P$ $^{a}$ (d)        & 3.7826330\\
T$_{0}$ (HJD)  & 2450003.58014 $\pm$ 0.01523\\
$e$            & 0.004 $\pm$ 0.001  \\
$\omega$ (deg)		&347 $\pm$ 3\\
V$\gamma$ (km/s)	&-28.12 $\pm$ 0.03\\
$K_1$ (km/s)		&82.08 $\pm$ 0.04\\
$K_2$ (km/s)		&110.08 $\pm$ 0.07\\
$M_1\sin ^3i$ ($M_\odot$)	&1.593 $\pm$ 0.002\\
$M_2\sin ^3i$ ($M_\odot$)	&1.188 $\pm$ 0.002\\
$q = M_2/M_1$		&0.7457 $\pm$ 0.0007\\
$a_1\sin i$ ($10^6$ km)	&4.2696 $\pm$ 0.0024\\
$a_2\sin i$ ($10^6$ km)	&5.7256 $\pm$ 0.0039\\
  \noalign{\smallskip}\hline
\end{tabular}
\end{center}
\end{table}

After the $v_{r}$ measurements, we derived the orbital parameters of DV\,Boo using the \texttt{rvfit} code\footnote{http://www.cefca.es/people/riglesias/rvfit html}. This code calculates the theoretical $v_{r}$ curves by using the adaptive 
simulated annealing method (\citealt{2015PASP..127..567I}). For the analysis, we took the orbital period (P$_{orb}$) and periastron passage time (T$_{0}$) 
from \cite{2004AcA....54..207K}. We also used the $v_{r}$ measurements given by \cite{2004MNRAS.352..708C}. The analysis was carried out using 74 $v_{r}$ measurements in total. During the analysis, only the P$_{orb}$ parameter was fixed and $e$, $w$, T$_{0}$, the velocity of the mass center V$\gamma$ and the amplitude of $v_{r}$ curve of the star relative to the center of the binary mass $K$ parameters 
were adjusted. The resulting parameters are given in Table\,\ref{table2} and 
the theoretical $v_{r}$ fit to the observed ones is shown in Fig.\,\ref{Fig1}.


\section{Spectral analysis}
\label{sect:analysis}
We aim to obtain the atmospheric parameters and the chemical composition of both binary component of DV\,Boo to examine their chemical structure and binary evolution. Therefore, in this section, we first disentangled the composite spectra of the system and obtained the individual spectra of binary components. Then, 
we carried out a detailed spectroscopic analysis using these disentangled spectra. 


\subsection{Spectral Disentangling}
\label{disen}
The spectral disentangling method can be used to acquire individual 
spectra of binary components in a double-lined (SB2) eclipsing binary system. 
To obtain the individual spectra of binary components the composite spectra of the binary system taken in different orbital phases and the fractional 
light contribution of binary components are needed. To obtain 
initial fractional lights of the binary components, we 
performed a preliminary light curve analysis using the ASAS data. In the initial light curve analysis, effective temperature 
($T_{\rm eff}$) value of the primary component was fixed. This $T_{\rm eff}$ value was estimated with the following steps. First, the Gaia distance (\citealt{2018A&A...616A...1G}) was used to estimate the interstellar absorption coefficient ($A{_v}$\,=\,0.062, \citealt{2011ApJ...737..103S}) and we calculated (B-V)${_0}$ value to be 0.31\,mag. By using the derived (B-V)${_0}$ and the list given by \cite{2018MNRAS.479.5491E} (see, Table\,7 in the article), we estimated the initial $T_{\rm eff}$ value of the primary binary component to be 7161\,K. As a result of the preliminary light curve analysis, we determined the 
light contributions to be around 80\% and 20 \% for primary and secondary components, respectively. 
The final light curve analysis will be given after the accurate $T_{\rm eff}$ values were obtained. 

The \texttt{FDBINARY} (\citealt{2004ASPC..318..111I}) code was used to obtain the individual spectra of the binary components. 
The \texttt{FDBINARY} 
disentangles the composite spectra basing on Fourier space by taking into account some orbital parameters 
such as $P$, T$_{0}$, $K_{1,2}$, $e$, and $\omega$. These orbital parameters can be fixed or 
adjusted during the analysis. However, the fractional light contributions of the binary components should be fixed in the analysis according to the orbital phases of the binary system in the used spectra.  

In the analysis, we used FEROS spectra because these data have significantly higher S/N ratio on average relative to HARPS and ELODIE data. Additionally, available FEROS spectra are also well distributed over the orbital phases of DV\,Boo for the disentangling analysis. 

In the analysis, the spectral range of around 4200\,$-$\,5600\,\AA\, 
was taken into account. This spectral window was divided into several spectral subsets with 150\,$-$\,250\,\AA\, steps. 
These spectral subsets were analysed separately. In the analysis, the input parameters were taken from the 
results of the radial velocity analysis and they were fixed. Only T$_{0}$ parameter was adjusted during the 
spectral disentangling process. After the individual spectra of binary components were obtained, these 
spectra were normalized considering the each component star' light fraction (\citealt{2005A&A...439..309P}). 

The S/N ratio of the resulting individual spectra of the binary components can be calculated using the given equation by 
\cite{2009MNRAS.394.1519P}. When we calculated the S/N ratios, we found that the spectra of 
primary and the secondary binary components have around 290 and 38 S/N ratio, respectively. 
The S/N ratio for the secondary component is low because of its less light contribution 
in total. Therefore, we are only able to determine the atmospheric parameters and metallicity value 
for the secondary star with this low S/N ratio. A detail abundance analysis will not be carried out for 
the secondary star.


  \begin{figure}
   \centering
   \includegraphics[width=10cm, angle=0]{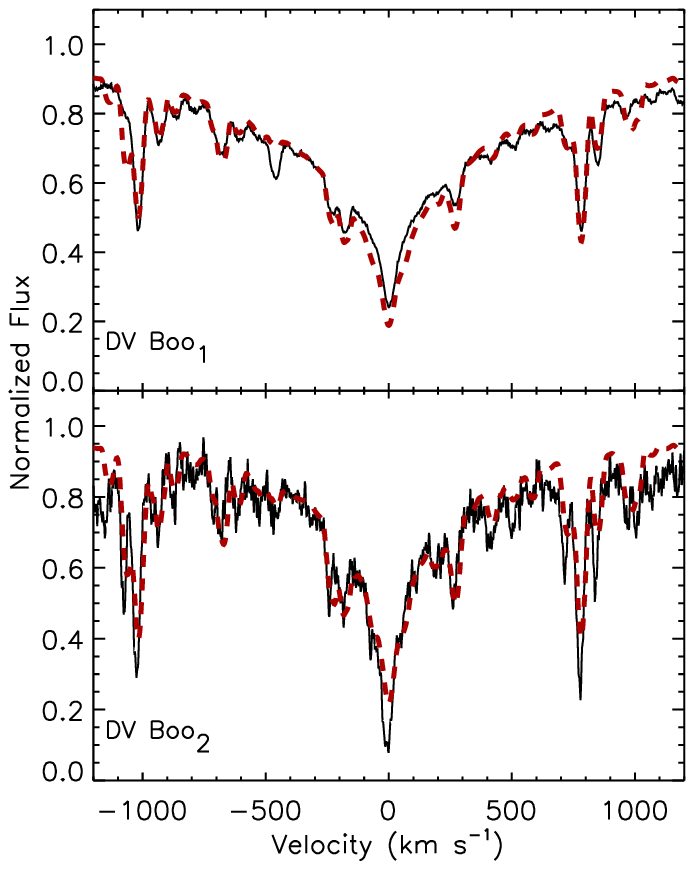}
   \caption{Theoretical fits (dashed lines) to the observed H$_{\gamma}$ lines. The subscripts ``1'' and ``2'' represent the primary and the secondary binary components, respectively.}
   \label{hlines}
   \end{figure}

\subsection{Determination of the atmospheric parameters and Abundance analysis}

To determine the atmospheric parameters ($T_{\rm eff}$, surface gravity $\log g$, 
micro-turbulence $\xi$) and projected rotational velocity ($v \sin i$) of both binary components, we used two different approaches. First, H$_{\gamma}$ line was taken into account to derive the $T_{\rm eff}$ parameter because hydrogen lines are very sensitive to $T_{\rm eff}$. It is also know that for cool stars ($T_{\rm eff}$\,$<$\,8000\,K) hydrogen lines are not sensitive to $\log g$ (\citealt{2002A&A...395..601S, 2005MSAIS...8..130S}). Hence, in the hydrogen line analysis, we took the $\log g$ to be 4.0\,cgs and additionally we assumed solar metallicity, as metallicity does not change the profile of hydrogen lines. The H$_{\gamma}$ $T_{\rm eff}$ values were derived considering the minimum difference between the synthetic and observed spectra. During this and future spectral analysis, the 
hydrostatic, plane-parallel and line-blanketed local thermodynamic equilibrium (LTE) ATLAS9 model atmospheres (\citealt{1993KurCD..13.....K}) were used. The synthetic spectra were generated by SYNTHE code (\citealt{1981SAOSR.391.....K}). 
 The errors in the H$_{\gamma}$ $T_{\rm eff}$ values were estimated considering the 1$\sigma$ difference in $\chi^2$ and also taken into account the possible uncertainty comes from the normalization ($\sim$100\,K Kahraman Alicavus, in preparation). The resulting H$_{\gamma}$ $T_{\rm eff}$ values are given in Table\,\ref{atmospar} and the theoretical fits to the H$_{\gamma}$ lines are shown in Fig.\,\ref{hlines}.

\begin{table}
\begin{center}
\caption[]{The Results of the Spectroscopic Analysis of the Individual Spectra of Binary Components. The Subscripts 1 and 2 Represent the Primary and the Secondary Binary Components, Respectively.}\label{atmospar}
 \begin{tabular}{lcccccc}
  \hline\noalign{\smallskip}
                 &\multicolumn{1}{c}{\hrulefill H$_{\gamma}$ line\,\hrulefill}
                 &\multicolumn{5}{c}{\hrulefill \,Fe lines\,\hrulefill}\\
Star             & $T_{\rm eff}$\,(K)   &  $T_{\rm eff}$\,(K)     & $\log g$\,(cgs)     & $\xi$\,(km\,s$^{-1}$)   & $v \sin i$\,(km\,s$^{-1}$) & $\log \epsilon$ (Fe)  \\
  \hline\noalign{\smallskip}
DV\,Boo$_{1}$   &7100\,$\pm$\,230   &7400\,$\pm$\,100   & 4.1\,$\pm$\,0.1    & 2.7\,$\pm$\,0.2 & 26\,$\pm$\,2      &7.79\,$\pm$\,0.25\\
DV\,Boo$_{2}$   &6500\,$\pm$\,360   &6500\,$\pm$\,100   & 4.3\,$\pm$\,0.2    & 3.7\,$\pm$\,0.3 & 17\,$\pm$\,3      &7.62\,$\pm$\,0.32\\
  \noalign{\smallskip}\hline
\end{tabular}
\end{center}
\end{table}

All atmospheric parameters were determined using another approach. In this approach, the excitation and ionization potential balances of iron (Fe) lines were taken into account. In this method, some small spectral parts ($1-5$\,\AA) were analysed separately by basing on the spectrum synthesis method. The final $T_{\rm eff}$, $\log g$, and $\xi$ parameters were searched in the range of $6000-7600$\,K, $3.8-4.5$\,cgs and $1-4$\,km\,s$^{-1}$ with 100\,K, 0.1\,cgs and 0.1km\,s$^{-1}$ steps, respectively. By using the given input parameters, we adjusted Fe abundances and $v \sin i$ parameters considering the minimum difference between the theoretical and observed spectra. The final $T_{\rm eff}$ and $\log g$ parameters were derived considering the relationships between the Fe abundance and excitation/ionization potentials. For the correct atmospheric parameters, these relationships should be flat because the abundance of the individual elements obtained from different lines should be the same for the different excitation potential. Additionally, $\xi$ value was obtained using the same abundance\,$-$\,exaction potential relationship as explained by \citealt{2014dapb.book..297S}. For the detailed explanation of the used method check the study of \citealt{2016MNRAS.458.2307K} and \citealt{2015MNRAS.450.2764N}.

In the analysis of primary component, 59 neutral and 24 ionized suitable Fe lines were used, while this number are 36 (FeI) and 15 (FeII) for the secondary component. The uncertainties in the determined final atmospheric parameters were obtained by the 1$\sigma$ change in the used relationships. We calculated how much the atmospheric parameters alter with the 1$\sigma$ difference in the considering relationships. The final atmospheric parameters and their uncertainties are given in Table\,\ref{atmospar}.

   \begin{figure}
   \centering
   \includegraphics[width=10cm, angle=0]{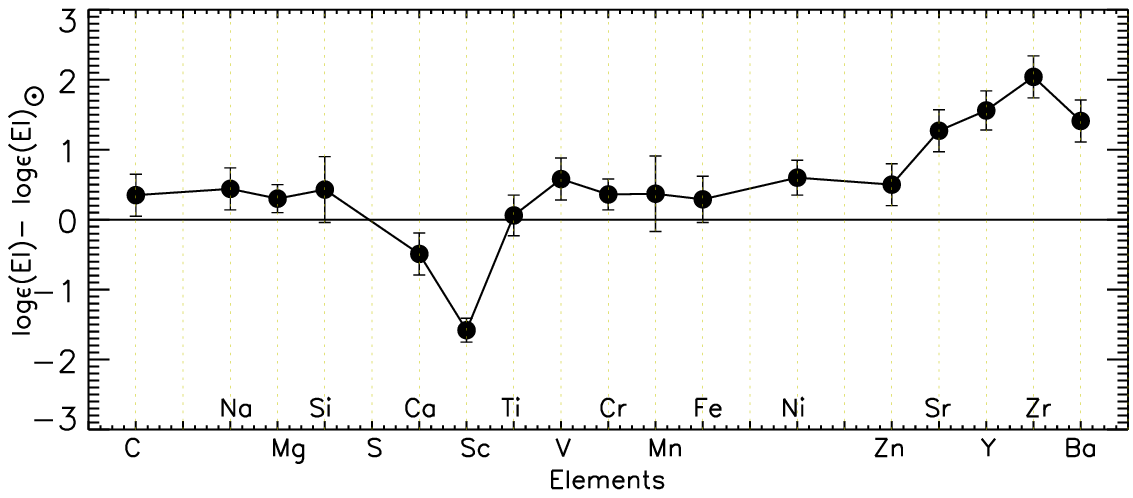}
   \caption{Chemical abundance distribution of the primary binary component relative to solar abundance (\citealt{2009ARA&A..47..481A})}
   \label{abundancedist}
   \end{figure}

   \begin{figure}
   \centering
   \includegraphics[width=12cm, angle=0]{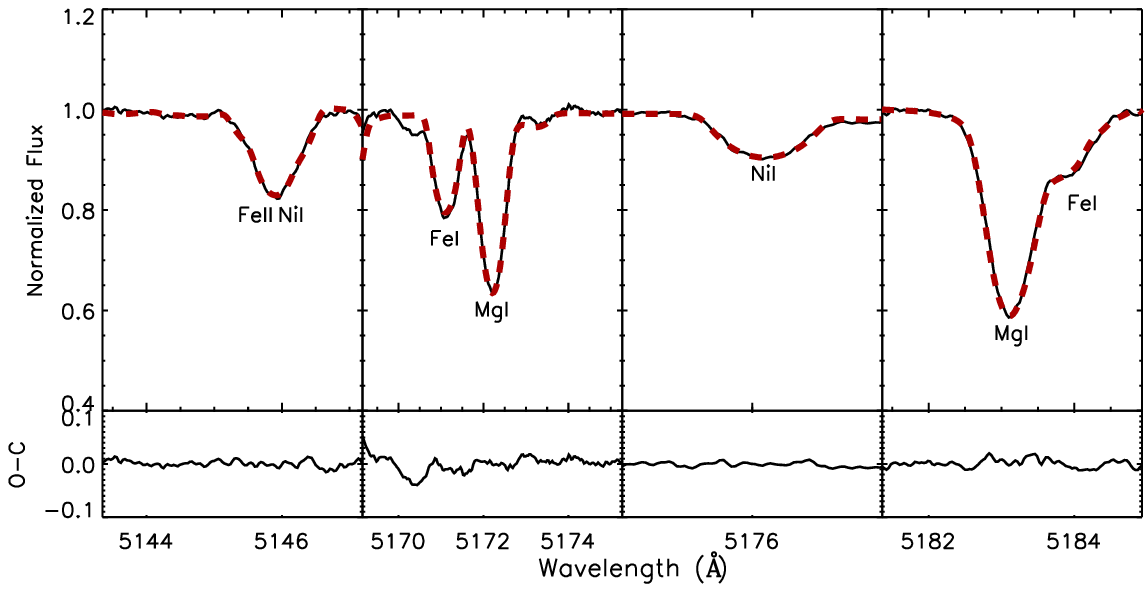}
   \caption{Theoretical fits (dashed lines) to the disentangled spectrum (solid lines) of the primary binary component (upper panels) and the residuals (O-C) (lower panels).}
   \label{abundancedist}
   \end{figure}

\begin{table}
\begin{center}
\caption[]{Abundances of Individual Elements of the Primary Star and Sun (\citealt{2009ARA&A..47..481A}). Number of the Analysed Spectral Parts is Given In the Brackets.}\label{abunresult}
 \begin{tabular}{lcc}
  \hline\noalign{\smallskip}
  Elements & Star abundance  & Solar abundance            \\
  \hline\noalign{\smallskip}
$_{6}$C   &8.78\,$\pm$\,0.44 (1)  & 8.43\,$\pm$\,0.05\\
$_{11}$Na &6.68\,$\pm$\,0.44 (1)  & 6.24\,$\pm$\,0.04                     \\
$_{12}$Mg &7.90\,$\pm$\,0.40 (4)  & 7.60\,$\pm$\,0.04\\
$_{14}$Si &7.94\,$\pm$\,0.39 (5)  & 7.51\,$\pm$\,0.03\\
$_{20}$Ca &5.85\,$\pm$\,0.41 (3) & 6.34\,$\pm$\,0.04\\
$_{21}$Sc &1.57\,$\pm$\,0.41 (3)  & 3.15\,$\pm$\,0.04\\
$_{22}$Ti &5.01\,$\pm$\,0.37 (15) & 4.95\,$\pm$\,0.05\\
$_{23}$V  &4.51\,$\pm$\,0.42 (2)  & 3.93\,$\pm$\,0.08			  \\
$_{24}$Cr &6.00\,$\pm$\,0.37 (15) & 5.64\,$\pm$\,0.04\\
$_{25}$Mn &5.80\,$\pm$\,0.39 (5)  & 5.43\,$\pm$\,0.05			   \\
$_{26}$Fe &7.79\,$\pm$\,0.25 (83) & 7.50\,$\pm$\,0.04\\
$_{28}$Ni &6.82\,$\pm$\,0.37 (19) & 6.22\,$\pm$\,0.04\\
$_{30}$Zn &5.06\,$\pm$\,0.44 (1)  & 4.56\,$\pm$\,0.05                     \\
$_{38}$Sr &4.14\,$\pm$\,0.45 (1)  & 2.87\,$\pm$\,0.07\\
$_{39}$Y &4.14\,$\pm$\,0.44 (1)  & 2.58\,$\pm$\,0.07\\
$_{40}$Zr &4.62\,$\pm$\,0.43 (2)  & 2.58\,$\pm$\,0.04			   \\
$_{56}$Ba &3.59\,$\pm$\,0.44 (1)  & 2.18\,$\pm$\,0.07\\
  \noalign{\smallskip}\hline
\end{tabular}
\end{center}  
\end{table}

 \begin{figure}
   \centering
   \includegraphics[width=12cm, angle=0]{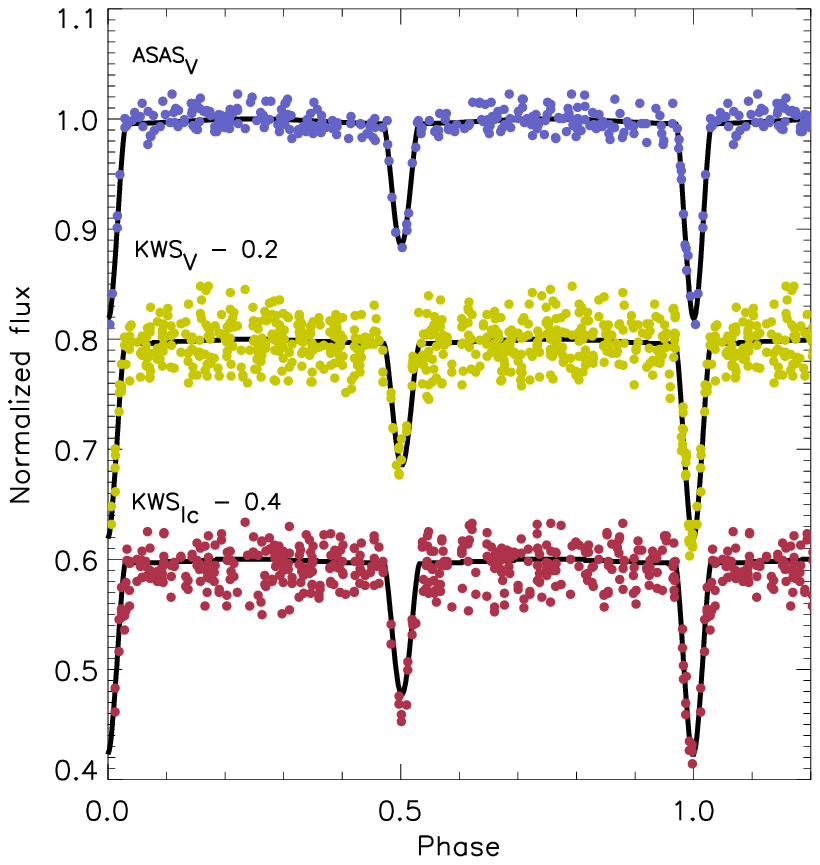}
   \caption{Theoretical light curve fits (solid lines) to the observed public photometric data (points) of DV\,Boo.}
   \label{lc}
   \end{figure}

After the final atmospheric parameters were determined, the abundance analysis was carried out taken these parameters as input. However, we only performed this analysis for the primary component, because the secondary component has a low S/N ratio for an abundance analysis. We only obtained $\log \epsilon$ (Fe) for the secondary component using the Fe lines because Fe lines are abundant and dominate in the secondary star's $T_{\rm eff}$range. For the abundance analysis of the primary component, first a line identification was done for each spectral part using the Kurucz line list \footnote{kurucz.harvard.edu/linelists.html} and then all parts were analyzed separately. The chemical abundances determined from the different spectral parts were taken into account to obtain the average individual chemical abundances. The list of final chemical abundances for the primary component is given in Table\,\ref{abunresult}. The uncertainties in the chemical abundances were estimated considering the effects of the quality of the spectrum (S/N, resolution), errors in the atmospheric parameters, and the assumptions in the model atmosphere calculations. The effect of local thermodynamical equilibrium (LTE) model assumption on chemical abundance calculations was searched by \cite{2011mast.conf..314M} and it turned out that these assumptions introduce around 0.1\,dex error. The effects of S/N and resolution were examined by \cite{2016MNRAS.458.2307K} 
in detail. We took into account the errors caused by these sources from this study. Additionally, we calculated the errors caused by the uncertainties 
in the determined atmospheric parameters. The quadrature sum of the errors introduced by the uncertainties of the atmospheric parameters was found to be around 0.15\,dex. The final calculated uncertainties are given in Table\,\ref{abunresult}.


\section{Light Curve analysis}

In the light curve analysis, the normalized $V$-band ASAS and $V$-, $Ic$-bands KWS photometric data were used. The analysis was carried out utilizing the Wilson-Devinney (\citealt{1971ApJ...166..605W}) code which was integrated with 
Monte Carlo (MC) simulation to precisely estimate the uncertainties in the adjusted parameters (\citealt{2004AcA....54..299Z, 2010MNRAS.408..464Z}).

\begin{table}
\begin{center}
\caption[]{Results of the Light Curve Analysis and the Astrophysical Parameters. The Subscripts 1, 2 and 3 Represent the Primary, the Secondary, and Third Binary Components, Respectively. $^a$ Shows the Fixed Parameters.}\label{lcresult}
 \begin{tabular}{lr}
  \hline\noalign{\smallskip}
   Parameter			  &  Value   	    \\
  \hline\noalign{\smallskip}
$i$ ($^{o}$)	       	          & 82.995 $\pm$ 0.251      \\	
$T$$_{1}$$^a$ (K)                 & 7400 $\pm$ 100  	   	\\	
$T$$_{2}$ (K)    	          & 6398 $\pm$ 174	    		\\
$V$$_{\gamma}$ (km/s)		  & -28.11 $\pm$ 0.03        	  \\
$a$ ($R$$_{\odot}$)        	  & 14.469 $\pm$ 0.010      	  \\
$e$$^a$	         	          &0.004$\pm$ 0.001        			  \\
$\Omega$$_{1}$		          & 8.100 $\pm$ 0.223      	  \\
$\Omega$$_{2}$		          & 9.499 $\pm$ 0.277     	  \\
Phase shift             	  & 0.0004 $\pm$ 0.0001   	  \\
$q$$^a$                     	  & 0.746 $\pm$ 0.001        	  \\
$r$$_{1}$$^*$ (mean)              & 0.1359 $\pm$ 0.0036      \\
$r$$_{2}$$^*$ (mean)              & 0.0889 $\pm$ 0.0026      \\
$L$$_{1}$ / ($L$$_{1}$+$L$$_{2}$) (ASAS) & 0.797 $\pm$ 0.016       	  \\
$L$$_{1}$ / ($L$$_{1}$+$L$$_{2}$) (KWS V) & 0.797 $\pm$ 0.016        	  \\
$L$$_{1}$ / ($L$$_{1}$+$L$$_{2}$) (KWS I) & 0.778 $\pm$ 0.016        	  \\
$L$$_{2}$ /($L$$_{1}$+$L$$_{2}$) (ASAS)  & 0.203 $\pm$ 0.016         \\
$L$$_{2}$ /($L$$_{1}$+$L$$_{2}$) (KWS V) & 0.203 $\pm$ 0.016      	  \\
$L$$_{2}$ /($L$$_{1}$+$L$$_{2}$) (KWS I) & 0.222 $\pm$ 0.016        	  \\
$l$$_{3}$                         & 0.0     			  \\
\multicolumn{2}{c}{Derived Quantities}\\

$M$$_{1}$ ($M_\odot$)	          & 1.629 $\pm$ 0.004  	    \\	
$M$$_{2}$ ($M_\odot$)	          & 1.215 $\pm$ 0.003       	\\
$R$$_{1}$ ($R_\odot$)	          & 1.966 $\pm$ 0.052      	  \\
$R$$_{2}$ ($R_\odot$)		  & 1.286 $\pm$ 0.038      	  \\
log ($L$$_{1}$/$L_\odot$)		  & 1.020 $\pm$ 0.033       		  \\
log ($L$$_{2}$/$L_\odot$)		  & 0.398 $\pm$ 0.054       		  \\
$\log g$$_{1}$ (cgs)               & 4.063 $\pm$ 0.023 	     			    \\
$\log g$$_{2}$ (cgs)               & 4.304 $\pm$ 0.025 	    			    \\
$M_{bolometric}$$_{1}$ (mag)      & 2.201 $\pm$ 0.082      		  \\
$M_{bolometric}$$_{2}$ (mag)	  & 3.755 $\pm$ 0.134       		  \\
$M_{V}$$_{1}$ (mag)	          & 2.166 $\pm$ 0.086      		  \\
$M_{V}$$_{2}$ (mag)	          & 3.759 $\pm$ 0.143       		  \\
Distance (pc)                     & 130  $\pm$ 5  \\           		
  \noalign{\smallskip}\hline
\end{tabular}
\end{center}
 \begin{description}
     \centering
 \item[ ] * fractional radii.
 \end{description}
\end{table}

During the analysis, some parameters were adjusted while some were fixed. The $T_{\rm eff}$ value of the primary star was set as 7400\,K from the present spectroscopic analysis. In addition to primary star's $T_{\rm eff}$ value, the logarithmic limb darkening coefficients taken from \cite{1993AJ....106.2096V} were fixed and also the bolometric albedos ($A$, \citealt{1969AcA....19..245R}), the bolometric gravity-darkening coefficients ($g$, \citealt{1924MNRAS..84..665V}) were fixed according to convective and radiative atmosphere assumption. We assumed these $A$ and $g$ values to be 1 for the primary component and respectively 0.5 and 0.32 for the secondary component. The parameters, $q$, $e$, and $\omega$ determined in the radial velocity analysis were also fixed. Additionally, the detached binary configuration was considered in the analysis. We adjusted the phase shift, $i$, secondary component's $T_{\rm eff}$, dimensionless potentials ($\Omega$) and the light fractions of  binary components. As a result, we obtained the parameters of the binary system. No third light contribution was found. The obtained values and their uncertainties calculated by MC method are given in Table\,\ref{lcresult}. The theoretical fits to the photometric data are presented in Fig.\,\ref{lc}. The fundamental stellar parameters were also calculated using the JKTABSDIM code (\citealt{2004MNRAS.351.1277S}). These parameters are also listed in Table\,\ref{lcresult}.


\section{Discussion and Conclusions}
\label{sect:discussion}
In this study, we present a detailed photometric and spectroscopic study of an eclipsing 
binary system DV\,Boo. The fundamental atmospheric parameters of binary components were obtained by 
using the disentangled spectra of each component. DV\,Boo was classified as a metallic-line (Am) star. 
According to our chemical abundance analysis, we found that the primary component demonstrates 
a typical Am star properties (see, Fig.\,\ref{abundancedist}). The primary star has mostly overabundant 
iron-peak elements and show deficiency in Ca and Sc elements. This is a typical chemical abundance 
behaviour of Am stars. It is known that many ($\sim$70\%) of Am stars are a member of binary systems (\citealt{2007MNRAS.380.1064C}). However, 
eclipsing binary Am stars are rare (\citealt{2014A&A...564A..69S}). Therefore, the current detailed analysis of DV\,Boo 
is important to deeply understand Am stars' behaviour. Additionally, we found that the secondary binary component also has a similar Fe abundance with the primary star.

When the obtained atmospheric parameters were examined, we noticed that the secondary binary component has a higher $\xi$ value comparing to the $\xi$ range for star with similar $T_{\rm eff}$ value (\citealt{2014psce.conf..193G}, \citealt{2009A&A...503..973L}). However, we keep in mind that our star is a member of a binary system and there are interactions between two binary components. Therefore, the reason of the resulting $\xi$ value could be the effect of binarity. This should be investigate in the further studies. 

 \begin{figure}
   \centering
   \includegraphics[width=12cm, angle=0]{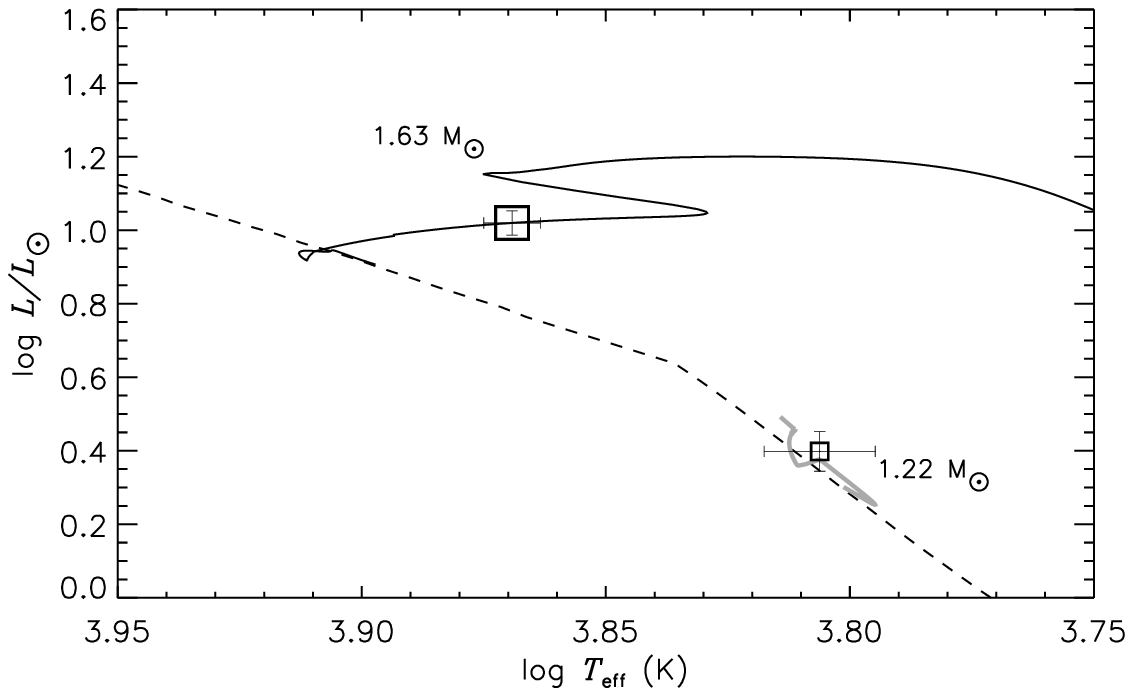}
   \caption{The positions of the primary and the secondary (smaller
symbol) binary components of DV\,Boo in the H-R diagram. Solid black and grey lines represent the evolutionary tracks for primary and secondary components, respectively. The dashed line illustrates the zero age main sequence (ZAMS).}
   \label{hr}
   \end{figure}
   
   
A light curve analysis was performed by using the ASAS V-band, KWS V- and Ic-bands data. As a result, we derived the orbital and the fundamental 
stellar parameters. The $M$ values of the binary components were obtained 
with an accuracy less than 1\%, while the accuracy in $R$ values is less 
than 3\%. These precise $M$ and $R$ values allow us to examine the evolutionary status of 
DV\,Boo. Therefore, the evolutionary status of DV\,Boo were examined by using the 8845\,version of the Modules for Experiments in Stellar Astrophysics (MESA) evolutionary programme (\citealt{2011ApJS..192....3P, 2013ApJS..208....4P}). The binary module of the MESA (\citealt{2015ApJS..220...15P}) was used to estimate the initial evolutionary parameters of the binary system 
and to model the orbital evolution. Many evolutionary models were generated with different input parameters. During the analysis, the metallicity ($Z$) value was searched between 0.01 and 0.02 with 0.001 steps. As a result, $Z$\,=\,0.015\,$\pm$\,0.002 was found. According to current spectroscopic analysis, the binary components of DV\,Boo have $Z$ value around solar ($Z$\,=\,0.0143, \citealt{2009ARA&A..47..481A}) within error bars. This $Z$ value is consistent with the $Z$ value obtained from the evolutionary models. Additionally, we estimated the initial orbital period and the $e$ value by comparing the models calculated with different input orbital parameters and $e$. 

\begin{figure}
   \centering
   \includegraphics[width=12cm, angle=0]{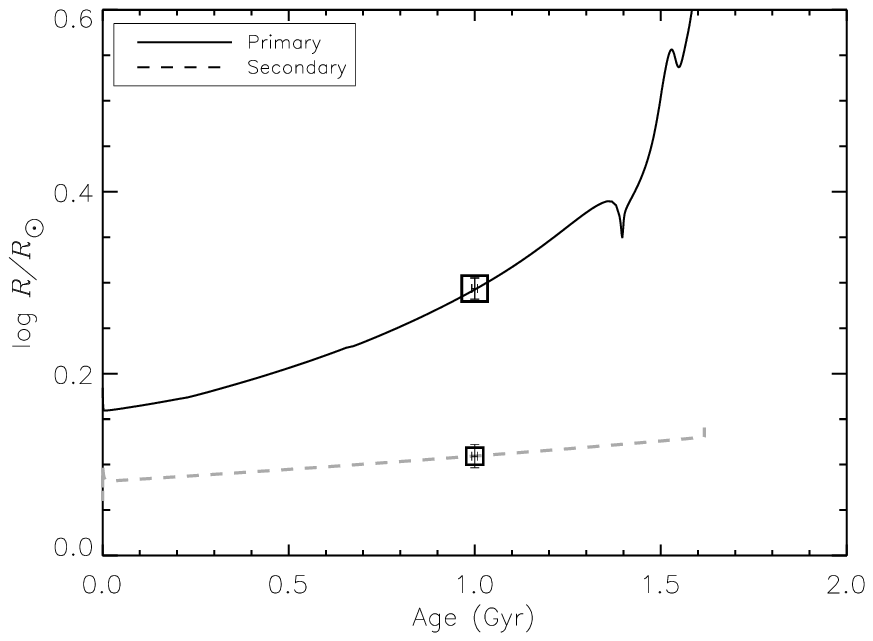}
   \caption{The position of the primary and the secondary (smaller
symbol) binary components of DV\,Boo in the Age\,$-$\,log\,$R$ diagram.}
   \label{age}
   \end{figure}

The resulting evolutionary models for primary and secondary binary components were estimated taking into account the best fit 
to the calculated Age\,$-$\,log\,$R$ diagram. As a result, we defined the age of components to be 1.00\,$\pm$\,0.08\,Gyr and examined the orbital evolution of the binary system. The first Roche lobe overflow (RLOF) time for the primary component of DV Boo is predicted to start at the age of 1.61 Gyr (after 0.61\,Gyr from the current age). The system is expected to become a semi-detached binary after this age with the beginning of rapid mass transfer. In this stage, the secondary component will gain mass and it will evolve parallel to the zero age main sequence (ZAMS) by its increasing radius and luminosity due to the mass transfer. The best 
fit evolutionary models for the binary components of DV\,Boo and the position of the components in the  Hertzsprung-Russell (H-R) diagram are shown in Fig.\,\ref{hr}. Furthermore, 
the location of the components in the Age\,$-$\,log\,$R$ diagram is illustrated in Fig.\,\ref{age}. In these figures, the binary components' evolutionary models are shown from the zero-age to the early stage of the beginning of the mass transfer. The estimated initial 
evolutionary parameters are given in Table\,\ref{evolres}.

\begin{table}
\begin{center}
\caption[]{Evolutionary Model Parameters Obtained from the MESA.}\label{evolres}
 \begin{tabular}{lc}
  \hline\noalign{\smallskip}
  Parameter                & Value       \\
                           &                  \\
  \hline\noalign{\smallskip}
$P$$_{initial}$  (days)    & 3.895\,$\pm$\,0.020    \\
$e$$_{initial}$            & 0.023\,$\pm$\,0.003          \\
Z                          & 0.015\,$\pm$\,0.002          \\
Age (Gyr)                  &1.00\,$\pm$\,0.08\\
  \noalign{\smallskip}\hline
\end{tabular}
\end{center}  
\end{table}

In the light curve analysis, a synchronous rotation was assumed. If the binary components are synchronised, their $v \sin i$ values should be 26.3\,$\pm$\,0.7 and 17.2\,$\pm$\,0.5 km\,s$^{-1}$ for the primary and the secondary components, respectively. These values are very similar to the spectroscopic ones. This shows us the binary components are already synchronised. In addition to this, using the resulting parameters of the light curve analysis we estimated the distance of DV\,Boo to be 130\,$\pm$\,5 pc which is consistent with the Gaia distance ($\sim$125\,pc, \citealt{2018A&A...616A...1G}).

DV\,Boo is given in the list of candidate $\delta$\,Scuti-type variables in eclipsing binaries 
(\citealt{2006MNRAS.370.2013S}). Therefore, we carried out a frequency analysis of the photometric data 
after removing the binary light variation. The Period04 code (\citealt{2005CoAst.146...53L}) was used in the analysis. 
We found some frequency peaks at $\delta$\,Scuti stars' pulsation frequency regime. However, the data 
is not good enough to classify the primary star to be $\delta$\,Scuti variable. A better quality data is needed.

The present detailed study of DV\,Boo offers good input data to examine the evolution of binary systems and to understand the characteristic of Am stars in binaries. This kind of spectroscopic analysis of eclipsing binary systems is required for a comprehensive investigation of binary evolution.  

\begin{acknowledgements}
The authors would like to thank the reviewer for his/her useful
comments and suggestions. FKA thanks the Polish National Center for Science (NCN) for 
supporting the study through grant 2015/18/A/ST9/00578.
We thank Prof. G. Handler for his helpful comments. 
The calculations have been carried out in Wroc{\l}aw Centre for Networking and Supercomputing (http://www.wcss.pl), grant No.\,214. 
This research has made use of the SIMBAD data base, operated at CDS, 
Strasbourq, France. This study is based on data obtained from the ESO Science Archive Facility under request number 524778 by Filiz Kahraman Alicavus and based on spectral data retrieved from the ELODIE archive at Observatoire de Haute-Provence (OHP, http://atlas.obs-hp.fr/elodie/). This work has made use of data from the European Space Agency (ESA) mission \emph{Gaia} (https://www.cosmos.esa.int/gaia), processed by the \emph{Gaia} Data Processing and Analysis Consortium (DPAC, https://www.cosmos.esa.int/web/gaia/dpac/consortium). Funding for the DPAC has been provided by national institutions, in particular the institutions participating in the \emph{Gaia} Multilateral Agreement.
\end{acknowledgements}



\appendix                  

\begin{table}
\begin{center}
\caption[]{The $v_{r}$ measurements. The subscripts ``1'' and ``2'' 
  represent the primary and the secondary components, respectively.}\label{rvs}
 \begin{tabular}{lccc}
  \hline\noalign{\smallskip}
HJD         &  $v_{r,1}$       &  $v_{r,2}$   &Instrument           \\
+2450000     &  (km\,s$^{-1}$)              &   (km\,s$^{-1}$)         \\
  \hline\noalign{\smallskip}
1931.6262     & -64.71\,$\pm$\,0.42  & 23.28\,$\pm$\,0.76  & ELODIE \\
2039.4253     & 9.91\,$\pm$\,0.25    & -77.80\,$\pm$\,0.86 & ELODIE \\
2040.5722     & -108.66\,$\pm$\,0.34 & 80.82\,$\pm$\,0.66  & ELODIE \\
2040.5722     & -109.26\,$\pm$\,0.48 & 82.58\,$\pm$\,0.98  & ELODIE \\
2041.4968     & -44.32\,$\pm$\,0.27  & -7.24\,$\pm$\,1.20  & ELODIE \\
2042.4086     & 52.12\,$\pm$\,0.37   & -136.04\,$\pm$\,0.91& ELODIE \\
2042.5900     & 53.97\,$\pm$\,0.37   & -137.16\,$\pm$\,1.02& ELODIE \\
2043.4398     & -20.74\,$\pm$\,0.61  &                     & ELODIE  \\
2297.6622     &-104.46\,$\pm$\,0.44  & 75.21\,$\pm$\,0.94  & ELODIE \\
2299.7025     & 54.31\,$\pm$\,0.29   &-137.86\,$\pm$\,0.71 & ELODIE \\
2303.7271     & 50.83\,$\pm$\,0.43   &-133.06\,$\pm$\,1.17 & ELODIE \\
2489.3476     & 33.27\,$\pm$\,0.31   &-107.19\,$\pm$\,0.84 & ELODIE \\
\hline
6473.5455     & -92.34\,$\pm$\,0.24  & 66.98\,$\pm$\,0.96  & FEROS \\
6473.6190     & -97.12\,$\pm$\,0.46  & 75.00\,$\pm$\,0.46  & FEROS \\
6474.5781     & -61.04\,$\pm$\,0.37  & 25.91\,$\pm$\,0.91  & FEROS \\
6474.6475     & -51.43\,$\pm$\,0.37  & 10.29\,$\pm$\,0.92  & FEROS \\
6475.5470     & 50.30\,$\pm$\,0.14   & -122.94\,$\pm$\,0.53& FEROS \\
6475.6228     & 54.41\,$\pm$\,0.24   & -128.43\,$\pm$\,0.84& FEROS \\
\hline
4887.8319     & -0.75\,$\pm$\,0.16   & -63.65\,$\pm$\,0.60 & HARPS \\
4887.8897     & -8.37\,$\pm$\,0.16   & -51.76\,$\pm$\,0.45 & HARPS \\
4889.7976     & -45.72\,$\pm$\,0.26  & -4.16\,$\pm$\,0.61  & HARPS \\
4890.8627     & 54.12\,$\pm$\,0.16   &-139.19\,$\pm$\,0.57 & HARPS \\
5431.4800     & 42.81\,$\pm$\,0.16   &-123.75\,$\pm$\,0.66 & HARPS \\
5432.4855     & -6.19\,$\pm$\,0.18   & -72.62\,$\pm$\,0.55 & HARPS \\
6449.5334     & 49.12\,$\pm$\,0.17   &-131.41\,$\pm$\,0.66 & HARPS \\
6450.5559     & -66.05\,$\pm$\,0.21  & -23.39\,$\pm$\,0.49 & HARPS \\
  \noalign{\smallskip}\hline
\end{tabular}
\end{center}
\end{table} 

\label{lastpage}

\end{document}